\documentclass[sigplan,9pt]{acmart}

\AtBeginDocument{%
  }

\setcctype[4.0]{by}
\setcopyright{cc}
\acmConference[TDIS '26]{4th International Workshop on Testing Distributed Internet of Things Systems}{April 27--30, 2026}{Edinburgh, Scotland Uk}
\acmBooktitle{4th International Workshop on Testing Distributed Internet of Things Systems (TDIS '26), April 27--30, 2026, Edinburgh, Scotland Uk}
\acmDOI{10.1145/3802513.3803486}
\acmISBN{}
\acmPrice{}
\acmYear{2026}
\copyrightyear{2026}

\usepackage{todonotes}
\usepackage{algorithm}
\usepackage{algpseudocode}
\usepackage{xspace}

\begin{document}

\newcommand{\newFramework}{EcoKube\xspace}
\newcommand{\newPolicy}{EcoKube Policy\xspace}
\newcommand{\policyLower}{ecokube-pol\xspace}

\title{EcoKube: Simulating Carbon-Aware Scheduling Policies in Heterogeneous Edge–Cloud Environments}

\author{Gonçalo Ferreira}
\email{g.j.teixeiradepinhoferreira@uva.nl}
\orcid{0000-0002-2819-1879}
\affiliation{%
  \institution{University of Amsterdam}
  \city{Amsterdam}
  \country{Netherlands}
}

\author{Shashikant Ilager}
\email{s.s.ilager@uva.nl}
\orcid{0000-0003-1178-6582}
\affiliation{%
  \institution{University of Amsterdam}
  \city{Amsterdam}
  \country{Netherlands}
}
\renewcommand{\shortauthors}{Ferreira et al.}

\begin{abstract}
    Energy demand from cloud and edge computing is rising rapidly, with AI workloads further intensifying electricity use and associated carbon emissions. 
    In hybrid edge--cloud settings, sustainability impact depends on time- and location-varying grid Carbon Intensity (CI), site Power Usage Effectiveness (PUE), and heterogeneous hardware characteristics. Existing carbon-aware work explores solutions such as temporal elasticity, spatio-temporal workload shifting, and carbon-aware placement across distributed sites. However, these solutions do not provide a consistent and reproducible workflow for evaluating sustainability-aware scheduling policies on heterogeneous, federated edge--cloud topologies. We present \newFramework: a configurable simulation framework for the reproducible evaluation of sustainability-aware scheduling policies in heterogeneous edge--cloud environments. The framework includes an event-driven deterministic simulator, policy hooks, and a heterogeneity-aware reference policy. We evaluate the framework with synthetic batch workloads, comparing the reference policy against the default Kubernetes scheduler, KEIDS, and TOPSIS/KCSS. The contribution is architectural and experimental: \newFramework provides a reproducible way to compare sustainability-aware policies before deployment.
\end{abstract}




\maketitle

\section{Introduction}
Global warming, largely driven by carbon dioxide emissions, remains one of the most pressing challenges of our time \cite{yangCarbonNeutralizedTaskScheduling2022}. At the same time, compute demand keeps ballooning: modern AI/ML training and inference, data-intensive science, and always-on edge services push more workloads across a widening cloud--edge continuum \cite{luccioniCountingCarbonSurvey2023}. This shift increases \textit{operational complexity}, as execution spans geo-distributed sites with different electricity mixes, facility overheads, and heterogeneous hardware, making sustainability impact more visible, but also more difficult to evaluate \cite{asadovCarbonAwareSpatioTemporalWorkload2025}.

A key approach for reducing operational footprint is through \textit{scheduling}: deciding \textit{where} and \textit{when} workloads run, and how resources are allocated during execution. However, assessing the sustainability impact of scheduling policies is challenging in hybrid edge--cloud settings because carbon- and energy-related signals are multi-level and time-varying: grid carbon intensity changes by region and time; site efficiency (e.g., PUE) shifts with load and season; and node-level performance-per-watt depends on hardware generation and accelerator availability \cite{yangSurveyTaskScheduling2025}.

Several state-of-the-art approaches explore carbon-aware workload management through temporal shifting (waiting for low-carbon windows), spatial shifting (moving computation to cleaner regions), or combinations of both \cite{wiesnerLetsWaitAwhile2021,radovanovicCarbonAwareComputingDatacenters2023,asadovCarbonAwareSpatioTemporalWorkload2025}. Kubernetes-native systems such as CarbonScaler demonstrate that forecast-driven, controller-based adaptation can improve carbon efficiency for temporally flexible workloads \cite{hanafyCarbonScalerLeveragingCloud2023}. Despite this progress, \textit{systematic evaluation under heterogeneity} remains challenging. Existing studies often focus on single-provider clusters, rely on ad-hoc experimental setups, or lack integrated modelling of both (i) \textit{site-level} signals (CI/PUE, effective CI) and (ii) \textit{node-level} constraints (hardware diversity, accelerators, interference), making results costly to reproduce and hard to compare across policies and topologies \cite{asadovCarbonAwareSpatioTemporalWorkload2025,yangSurveyTaskScheduling2025,ghafariTaskSchedulingAlgorithms2022}. What is needed is a testable workflow that supports controlled, repeatable experiments across heterogeneous, federated edge--cloud scenarios while remaining close to Kubernetes operational realities \cite{senjabSurveyKubernetesScheduling2023}.

In this paper, we present \textit{\newFramework}, a configurable framework for testing sustainability-aware scheduling policies in heterogeneous edge--cloud environments. \newFramework provides a discrete-event simulation workflow that (i) models site-level sustainability signals (CI and site normalisation), (ii) captures node-level heterogeneity and feasibility constraints (latency and accelerator fit), and (iii) exposes transparent policy hooks to implement and compare practical heuristics and multi-criteria schedulers \cite{menouerKCSSKubernetesContainer2021,kaurKEIDSKubernetesBasedEnergy2020,senjabSurveyKubernetesScheduling2023}. Concretely, \newFramework contributes: (1) a deterministic modular simulator for hybrid edge--cloud scheduling experiments; (2) a reference policy instantiation, \textit{\newPolicy}, that demonstrates how site-level sustainability signals and node-level feasibility constraints can be composed within the framework; and (3) an evaluation methodology for comparing sustainability--performance trade-offs across scenario families \cite{asadovCarbonAwareSpatioTemporalWorkload2025,ghafariTaskSchedulingAlgorithms2022}.

Our experiments show that explicitly modelling heterogeneity changes the observed trade-offs between carbon and energy and performance, and that sustainability-aware policies can reduce the emissions estimate under the evaluated hybrid scenarios without requiring impractical assumptions about uniform infrastructure or perfect testbed control. In this evaluation, \newPolicy serves as a transparent reference instantiation used to validate the framework rather than as a claim of a new optimisation paradigm.
This behaviour is consistent with prior evidence that carbon-aware orchestration can reduce emissions under temporal flexibility and operational constraints \cite{wiesnerLetsWaitAwhile2021,hanafyCarbonScalerLeveragingCloud2023}. The remainder of the paper is organised as follows: \S\ref{sec:background_related} summarises the required background and related work; \S\ref{sec:solution_overview} describes the \newFramework simulator, policy interface and implementation; \S\ref{sec:performance_evaluation} describes the setup of the experiment and evaluates its performance; and \S\ref{sec:conclusions_future} concludes with limitations and future work.

\section{Background and Related Work}
\label{sec:background_related}

Hybrid edge--cloud systems span multiple \textit{sites} (regions, edge clusters, or federated Research Infrastructure (RI) facilities) that differ in hardware capabilities, operational efficiency, and carbon conditions. This heterogeneity matters because the \textit{same} workload can incur materially different energy use, completion time, and carbon footprint depending on where it runs \cite{radovanovicCarbonAwareComputingDatacenters2023}. Consequently, sustainability-aware scheduling is typically framed as a multi-objective optimisation problem that balances carbon impact against latency- and makespan-oriented SLOs \cite{hongResourceManagementFog2020}.

\paragraph{Carbon-intensity, site overhead, and normalisation.}
Carbon-aware orchestration relies on exogenous signals: time-varying grid carbon intensity $\mathrm{CI}_s(t)$ (e.g., gCO$_2$e/kWh) for each site $s$, and facility efficiency captured via Power Usage Effectiveness $\mathrm{PUE}_s$ \cite{radovanovicCarbonAwareComputingDatacenters2023,katalEnergyEfficiencyCloud2023}. A common abstraction is to express job-level carbon as a function of IT energy, site overheads, and carbon intensity:
\begin{equation}
\mathrm{CFP}(j,s) \approx E^{\mathrm{IT}}_{j,s} \cdot \mathrm{PUE}_s \cdot k_s \cdot \overline{\mathrm{CI}}_s,
\label{eq:cfp_simple}
\end{equation}
where $E^{\mathrm{IT}}_{j,s}$ is the workload IT energy attributable to job $j$ at site $s$, $\overline{\mathrm{CI}}_s$ is the average carbon intensity over the job interval, and $k_s$ is a site normalisation factor that compensates for cross-site measurement and attribution differences in heterogeneous environments \cite{saadCarbonAwareContainerOrchestration2025}. While many studies treat $\mathrm{CI}_s(t)$ as a given external input, hybrid and federated settings also motivate \textit{effective} CI models (e.g., accounting for imports/exports, local generation mix, or finer-grained spatial variation), which remain inconsistently considered in scheduling-oriented frameworks \cite{wuCarbonEdgeLeveragingMesoscale2025}.

\paragraph{Latency modelling and accelerator fit.}
Beyond sustainability signals, edge--cloud scheduling is constrained by responsiveness and feasibility. We model end-to-end completion time as the sum of dispatch/queue delay and execution time, enriched with network terms:
\begin{equation}
T_j = q_j + r_{j,s,n} + \ell^{\mathrm{net}}_{s}(j),
\label{eq:latency_model}
\end{equation}
where $r_{j,s,n}$ depends on both site $s$ and node $n$ due to performance heterogeneity and contention effects \cite{hongResourceManagementFog2020}. \textit{Node/accelerator fit} further constrains feasible placement, capturing whether a node's resources (e.g., GPU class, memory headroom) match a workload's typology requirements \cite{kaurKEIDSKubernetesBasedEnergy2020}.

\paragraph{Frameworks and algorithmic baselines.}
Prior sustainability frameworks demonstrate strong results for specific levers such as temporal adaptation, energy-system virtualisation, job-level optimisation, and geo-distributed carbon-aware placement/provisioning \cite{saadCarbonAwareContainerOrchestration2025,souzaCASPERCarbonAwareScheduling2024}. In parallel, multi-criteria decision methods (e.g., TOPSIS-style ranking) and edge-focused schedulers (e.g., KEIDS-like multi-objective policies) provide lightweight online baselines that combine energy/carbon signals with performance and contention considerations \cite{menouerKCSSKubernetesContainer2021,kaurKEIDSKubernetesBasedEnergy2020}. However, existing approaches rarely provide a \textit{unified, testing-oriented} workflow that (i) makes site-level placement decisions using carbon and efficiency signals, while also (ii) modelling node-level heterogeneity (including accelerators) and (iii) enabling repeatable comparison across heterogeneous, federated edge--cloud topologies \cite{saadCarbonAwareContainerOrchestration2025}. 
This gap motivates \newFramework as a reproducible, testing-oriented workflow for hybrid settings. Within that workflow, \newPolicy is used as a reference policy instantiation that leverages the separation of site-level signals and node-level constraints.

Exploring this gap, our objective is a neutral and reproducible workflow for \textit{comparing} Kubernetes-compatible and sustainability-aware policies under identical heterogeneous conditions, including site-level CI/PUE differences and node-level feasibility constraints. This motivates \newFramework as a testing-oriented simulator built from scratch for hybrid edge--cloud settings. Within that workflow, \newPolicy is used only as a transparent reference instantiation.

\section{\newFramework: Simulating Carbon-Aware Scheduling Policies in Heterogeneous Edge-Cloud Systems}
\label{sec:solution_overview}

\begin{figure*}[t]
  \centering
  \includegraphics[width=0.8\textwidth]{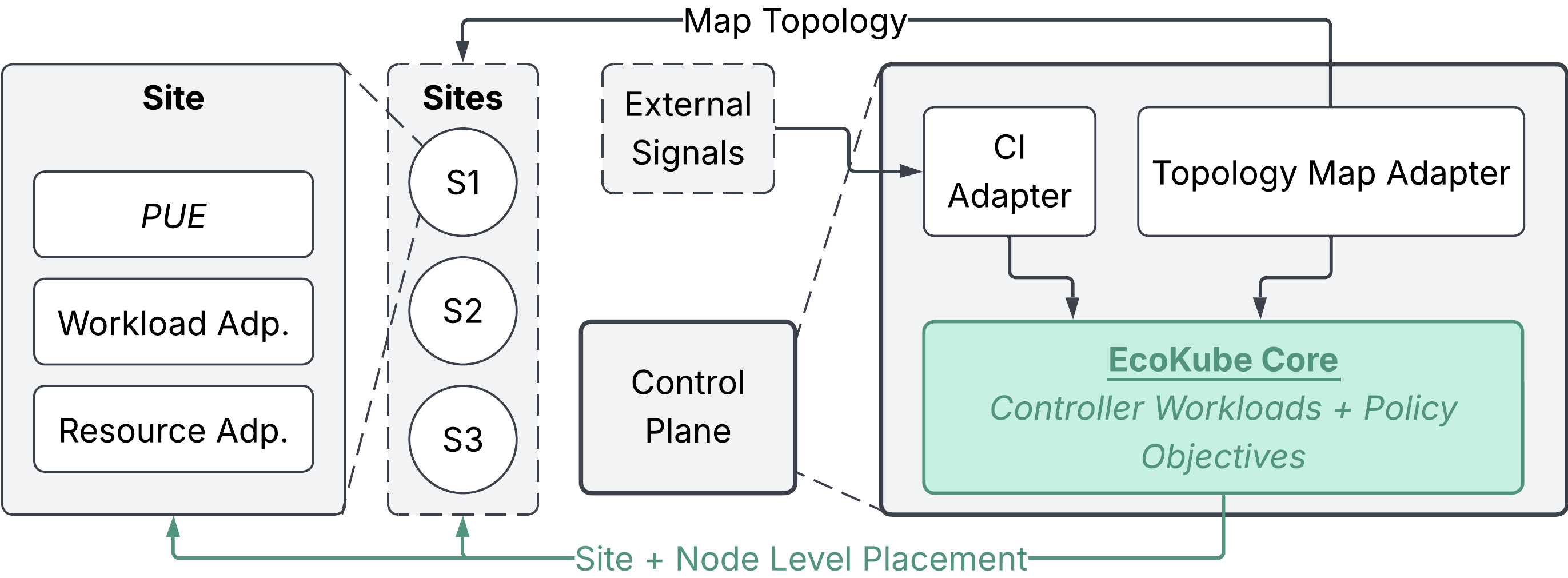}
  \caption{EcoKube System Model}
  \label{fig:system_model}
\end{figure*}

In this section we explain how \newFramework operationalises sustainabi-lity-aware scheduling in a heterogeneous, multi-site setting. We first define a \textit{system model} (Figure \ref{fig:system_model}) that captures the real-world actors, signals, and control loops required to define workload placement using sustainability signals. We then show how this model is \textit{abstracted and executed in simulation}, preserving the same interfaces and decision points while enabling controlled experimentation. Finally, we describe the \textit{experiment configuration} used to instantiate the model (workload traces, site/topology parameters, and policy settings) and to generate the results reported afterwards.

\subsection{System Model}
\label{sec:system_model}
\newFramework is designed for \textit{edge--cloud} deployments, where sites differ not only in hardware, but also in \textit{geography, network connectivity, and computing capability}. Edge sites typically provide proximity and reduced data movement, but operate under tighter capacity constraints and higher variability (e.g., intermittent availability, limited accelerators, and constrained cooling). Cloud and core data-centre sites offer higher and more elastic capacity, but may incur higher data transfer overheads and different sustainability profiles. This inherent heterogeneity motivates a placement policy that trades off \textit{carbon-, energy-, and performance-related} objectives across the edge--cloud continuum rather than assuming a uniform cluster environment.

\subsubsection{Site Characteristics}
\label{sec:system_model_actors}

Each site exposes a set of hooks necessary for sustainability-aware placement:
\begin{enumerate}
    \item \textbf{PUE descriptor:} a site-level efficiency factor that approximates facility overheads relative to IT power.
    \item \textbf{Resource Adapter:} a site-side interface that exposes \textit{capacity} and \textit{capabilities} (e.g., CPU/memory/GPU availability, accelerator types); as well as static constraints relevant to feasibility (e.g., resource limits, hardware incompatibilities).
    \item \textbf{Workload Adapter:} an interface that describes workload requirements in a portable form (e.g., requested resources, runtime class, optional locality constraints).
\end{enumerate}

In our prototype, workloads execute on a federated multi-site testbed where each site exposes these local characteristics (capacity/capabilities and PUE) and is enriched with external, time-varying signals (e.g., CI). These inputs are ingested and normalised by the metrics and KPI services and consumed by the EcoKube's control plane, which applies policies (via the Resource/Workload adapters) to select feasible sites/nodes and drive placement decisions in the orchestrator engine.

\subsubsection{\newFramework Components}
\label{sec:system_model_components}

Within the \newFramework boundary (Figure~\ref{fig:system_model}), the model distinguishes three components.

\begin{enumerate}
    \item \textbf{CI Adapter:}
    The CI adapter retrieves CI for the relevant scope (e.g., region, zone, or site mapping), and it wires it into the policy. In the real system, this component is responsible for handling harmonisation issues (e.g., sampling frequency, caching, missing values, and time alignment). In the simulation, the same interface is used but backed by trace-driven or synthetic CI series (Section~\ref{sec:experimental_setup}).
    
    \item \textbf{Topology Map Adapter:}
    This adapter provides the mapping between workloads, sites, and topology constraints. Minimally, it maps \textit{site identifiers} to \textit{signal scope identifiers} (e.g., which CI region applies to which site). In more advanced scenarios it can express locality constraints or penalties (e.g., prefer a site due to data proximity; penalise sites due to cross-site transfer).
    
    \item \textbf{Core: Controller Workloads + Policy Objectives:}
    In particular for edge--cloud scenarios, the topology map captures \textit{cross-site effects} such as latency, bandwidth, and data gravity (where datasets or users are located). These signals allow \newFramework to represent the cost of placing an otherwise ``green'' workload in terms of its ``network cost''.
\end{enumerate}

\subsubsection{EcoKube Scoring}
EcoKube makes decisions in two stages. It first filters infeasible sites and nodes using standard constraints such as resource availability, placement restrictions, and locality requirements. It then ranks the remaining candidates using a small set of normalised signals: estimated IT energy, a location-aware estimated emissions term derived from energy, PUE, and CI, latency-related penalties, and hardware fit.

\paragraph{Site- and node-level scoring.}
Within the framework, \newPolicy decomposes each decision into (i) a \textit{site-level} score capturing exogenous sustainability and topology signals and (ii) a \textit{node-level} score capturing energy, latency, and hardware fit. At the site level, candidates are ranked using normalised site descriptors, namely CI, PUE, a site normalisation factor, and an optional network penalty when topology-aware routing is enabled. These terms are combined through configurable weights to obtain a coarse inter-site score.

Within the selected site, each feasible node $n$ is scored as:
\begin{equation}
\label{eq:node-cost}
\begin{aligned}
C(w,n,t)=\;&
w_E\,{\widehat{E}_{w,n}}
+ w_C\,{\widehat{E}_{w,n}\cdot \overline{\mathrm{PUE}}_{s(n)}\cdot \overline{\mathrm{CI}}_{s(n)}}\\
&+ w_L\,{\mathrm{lat}(w,n)}
+ w_{fit}\,\bigl(1-{\mathrm{fit}(w,n)}\bigr),
\end{aligned}
\end{equation}
where $\widehat{E}_{w,n}$ is the estimated IT energy for placing workload $w$ on node $n$, $\mathrm{lat}(w,n)$ is the latency-related penalty, and $\mathrm{fit}(w,n)$ captures hardware suitability. Lower scores are preferred. The weight vectors are treated as configuration parameters for experimentation rather than as universally optimal coefficients. The site-level score and node-level score play different roles in the decision process: the former captures coarse inter-site context, while the latter ranks feasible nodes within the selected site using workload-specific estimates. Each component is min–max normalised over the feasible candidate set.

\paragraph{Selection Rule.}
The final decision combines the exogenous site score with the best feasible node score within each site. Intuitively, the policy first prefers \textit{cleaner and more efficient} sites (CI/PUE/$k$, optionally network penalties), and then selects the \textit{most suitable} node at that site according to energy, latency and accelerator fit.

\begin{equation}
\label{eq:select-min}
(s^\star,n^\star)=\arg\min_{s}\Bigl(C_{\mathrm{site}}(w,s,t)+\min_{n\in s} C(w,n,t)\Bigr).
\end{equation}

Here, \(C_{\mathrm{site}}(w,s,t)\) is the site-level score for workload \(w\) at site \(s\) and time \(t\), \(C(w,n,t)\) is the node-level score for candidate node \(n\), and the inner minimisation selects the best feasible node within each site before comparing sites globally.
\newPolicy instantiates the framework through a weighted ranking over feasible candidates. This reference policy is used to demonstrate how the framework can combine heterogeneous inputs such as site-level PUE/CI descriptors, normalisation factors, and node-level fit and latency signals within one reproducible workflow. This also makes the policy suitable for controlled comparison against existing schedulers and for sensitivity analysis over configuration weights.

The weight vectors are therefore treated as experimental configuration knobs. We sweep them to study policy behaviour and robustness, rather than to argue for a single theoretically optimal setting. The default values reported in the evaluation were selected as a reasonable reference configuration after preliminary tuning; they are intended to provide a stable comparison point rather than to imply Pareto-optimality or theoretical optimality.

\subsubsection{EcoKube Configuration: Policy Weights}
\label{sec:policy_weights}
For each feasible candidate, EcoKube computes comparable normalised terms for energy, emissions, latency, and fit, and combines them through non-negative weights that sum to one; the hardware-fit term is controlled separately through \(w_{fit}\) and is not included in that normalisation constraint. Lower scores are preferred. The estimated emissions term is an operational estimate obtained by combining expected IT energy with the candidate site's PUE and time-aligned grid carbon intensity. Energy and emissions are kept as distinct signals on purpose: energy captures hardware efficiency, whereas the emissions term captures the execution context of the selected site. The weights are configuration parameters used to express operator preferences in the experiments; they are not claimed to be universally optimal.

\newPolicy is parametrised by a weight vector $W=\langle w_E,w_C,$ $w_L,w_{fit}\rangle$, where each component controls the relative importance of estimated IT energy, estimated emissions, latency penalties, and node/\allowbreak accelerator fit in Eq.~\ref{eq:node-cost}, respectively. In addition, the policy uses separate site-level configuration parameters to control the contribution of site-level sustainability and placement terms during candidate-site ranking. In \newFramework, these weights are treated as exposed configuration parameters that instantiate different operational priorities within the same evaluation workflow, rather than as globally optimal constants. Their role is therefore experimental and comparative: they make policy behaviour explicit and sweepable without changing the underlying decision logic.

\subsection{Implementation}
\label{sec:implementation}
\newFramework was implemented from scratch as a methodological choice to support controlled and reproducible evaluation across heterogeneous federated edge--cloud scenarios. Extending an existing scheduler or controller would have made the evaluation dependent on a specific system's execution model, making controlled comparison across heterogeneous federated edge--cloud scenarios more difficult. At the same time, \newFramework remains connected to Kubernetes by preserving Kubernetes-style feasibility filtering and pluggable scoring logic, so it can be used to evaluate improvements on top of Kubernetes-compatible scheduling workflows rather than as a replacement for them.

\newFramework is therefore implemented in \textit{Go (Golang)} as a modular scheduling engine that separates the \textit{core loop} (queueing, feasibility filtering, tie-breaking, and binding) from the \textit{policy logic} (feature extraction and scoring). The scheduler constructs a feasible candidate set using Kubernetes-style hard constraints (requests/limits, taints/tolerations, affinity), then delegates ranking to a pluggable \texttt{Score} hook exposed by each policy module. This design allows \newFramework and all baselines to reuse identical inputs (site descriptors, CI/PUE signals, and node capabilities) and to be swapped without changing the execution pipeline, enabling like-for-like comparisons under controlled scenarios. We use Python scripts for post-processing and plotting the experimental results.

Because feasibility filtering follows Kubernetes-style constraints and ranking is exposed through a pluggable \texttt{Score} hook, the framework is intended to improve policy design \textit{on top of} Kubernetes-like schedulers, not replace them. Policies can first be evaluated in \newFramework under controlled heterogeneity and then translated into Kubernetes-compatible extensions for deployment validation.

Carbon intensity traces are sourced from the Wattnet service, using 2024 time-series data retrieved through their API\footnote{\url{https://api.wattnet.eu}}. These traces are ingested by the CI adapter and aligned to scenario time windows to provide consistent, time-varying exogenous signals across all policies and repetitions.

\section{Performance Evaluation}
\label{sec:performance_evaluation}

We evaluate \newFramework under a scenario-driven pipeline designed to reflect \textit{hybrid edge--cloud} placement decisions, where sites differ in capacity, device heterogeneity, topology constraints, and time-varying sustainability signals. To ensure fair comparisons, all schedulers are executed on identical scenario instances (same topology, traces, feasibility constraints), using fixed random seeds and matched repetitions. All experimental artifacts (code, scenario configurations, and seeds) are available online on GitHub.\footnote{\url{https://github.com/g-uva/EcoKube/tree/tdis-26}}

\subsection{Experimental Setup}
\label{sec:experimental_setup}

\paragraph{Scenarios, topology, and workloads.}
We evaluate \newFramework on a deterministic three-site edge--cloud topology spanning \texttt{NL}, \texttt{FR}, and \texttt{DE}. The generated substrate contains eight heterogeneous nodes: two small CPU-oriented nodes in \texttt{NL}, two balanced CPU nodes and one GPU-capable node in \texttt{FR}, and two memory-oriented nodes plus one GPU-capable node in \texttt{DE}. Site descriptors are fixed per campaign and include Power Usage Effectiveness (PUE), a site normalisation factor $k_s$, and hourly carbon-intensity traces.
Workloads are produced by a single parameterised generator and instantiated as three workload families. The preset name therefore denotes the workload family and heterogeneity pressure; submit timestamps are regenerated by the reported arrival-process sweep so that all policies observe identical arrivals per scenario key.

\begin{table}[t]
\centering
\small
\caption{Workload presets used in the simulator evaluation.}
\label{tab:scenarios_workloads}
\begin{tabular}{p{0.18\linewidth} p{0.74\linewidth}}
\toprule
\textit{Preset} & \textit{Description} \\
\midrule
\texttt{mix-a} & CPU-centric monitoring mix with low accelerator demand (\texttt{gpu\_share} $=0.05$), intended to represent stable edge sensing and lightweight processing pipelines. \\
\texttt{mix-b} & Mixed edge-event workload with moderate accelerator demand (\texttt{gpu\_share} $=0.12$), intended to represent intermittently heavier event-driven processing. \\
\texttt{mix-c} & Stronger heterogeneous CPU/GPU mix (\texttt{gpu\_share} $=0.28$), intended to stress feasibility filtering, device affinity, and node-fit decisions. \\
\bottomrule
\end{tabular}
\end{table}

\begin{figure*}[t]
  \centering
  \includegraphics[width=\textwidth,]
  {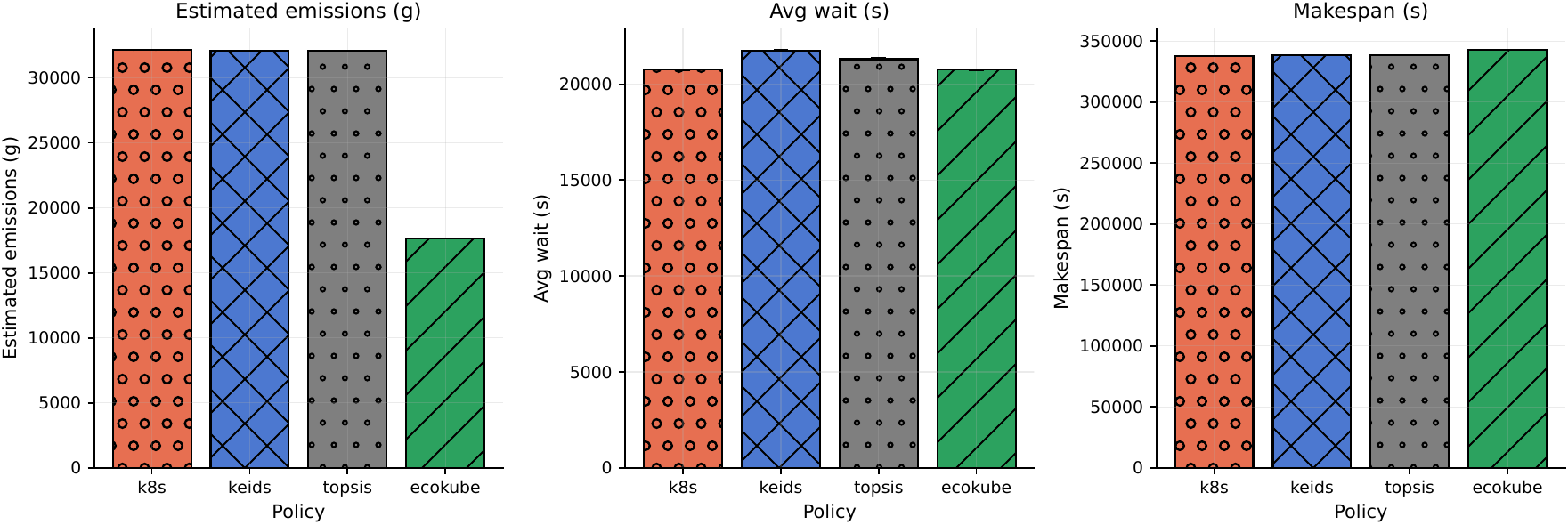}
  \caption{Policy Outcome Comparison}
  \label{fig:policy-comparison}
\end{figure*}

\paragraph{Baselines.}
We compare \newPolicy against a default Kubernetes baseline and three policy baselines commonly used to capture multi-objective ranking and carbon-aware placement behaviours. Each baseline uses the same feasibility constraints and observes the same external signals (Table \ref{tab:baselines}).

\begin{table}[t]
\centering
\small
\caption{Schedulers compared in the evaluation.}
\label{tab:baselines}
\begin{tabular}{p{0.22\linewidth} p{0.70\linewidth}}
\toprule
\textit{Policy} & \textit{Role in comparison} \\
\midrule
\texttt{k8s} & Default baseline (reference for deltas). \\
\texttt{keids} \cite{kaurKEIDSKubernetesBasedEnergy2020} & Heuristic sustainability-aware baseline (CI-sensitive ranking). \\
\texttt{topsis} \cite{menouerKCSSKubernetesContainer2021} & Multi-criteria decision baseline using ranking over normalised objectives. \\
\texttt{\policyLower} & Reference policy instantiation in \newFramework: weighted-sum scoring over estimated emissions term, performance, topology penalties, and device-fit. \\
\bottomrule
\end{tabular}
\end{table}

\paragraph{Reproducibility and parameter sweeps.}
All simulator runs are deterministic per scenario key. We fix the workload-generation seed, the arrival-schedule seed, and the policy execution order, and reuse identical generated inputs across all policies. The reported evaluation varies the carbon-intensity weight $\theta_c$, the arrival rate, the batch size, and the number of jobs, while keeping the topology and site descriptors fixed. A 30-minute warm-up window is excluded from summary metrics.

\begin{table}[t]
\centering
\small
\caption{Reproducibility controls and sweep parameters used in the simulator campaign.}
\label{tab:sweeps}
\begin{tabular}{p{0.40\linewidth} p{0.52\linewidth}}
\toprule
\textit{Parameter / control} & \textit{Values} \\
\midrule
Sites & $\{\texttt{NL}, \texttt{FR}, \texttt{DE}\}$ \\
Generation seed & \texttt{20260214} \\
Job counts & $\{300, 600, 900\}$ \\
Arrival rates (jobs/min) & $\{0.8, 1.1, 1.4\}$ \\
Batch sizes & $\{200, 500, 900\}$ \\
Carbon weight sweep (
$\theta_c$) & $\{0.2, 0.4, 0.6, 0.8\}$ \\
Warm-up window & $30$ minutes \\
Repetitions per scenario & $50$ \\
Site-level parameters & $\alpha=0.58,\; \beta=0.21,\; \gamma=0.21,\; w_{\textit{fit(s)}}=0.2$ \\
Node-level weights & \textit{$w_E=0.58,\quad w_C=0.21,\quad w_L=0.21,\quad w_{\textit{fit(n)}}=0.20$} \\
\bottomrule
\end{tabular}
\end{table}

\begin{table*}[t]
\centering
\small
\caption{Absolute outcomes for the reported evaluation slice.}
\label{tab:absolute_outcomes}
\begin{tabular}{l r r r r r}
\toprule
\textit{Policy} & \textit{Est. Emissions (kg)} & \textit{Avg.\ Est. Emissions/job (g)} & \textit{Makespan (s)} & \textit{Avg.\ wait (s)} & \textit{Completed jobs} \\
\midrule
\texttt{\policyLower} & 17.63 & 6.63 & 342600 & 20749 & 848 \\
\texttt{topsis}       & 32.12 & 15.17 & 338259 & 21298 & 849 \\
\texttt{keids}        & 32.12 & 15.63 & 338329 & 21748 & 842 \\
\texttt{k8s}          & 32.14 & 15.40 & 338017 & 20758 & 848 \\
\bottomrule
\end{tabular}
\end{table*}

\paragraph{Evaluation metrics and interpretation.}
We report three classes of outcomes. First, we report scheduler-external metrics, namely estimated IT energy, makespan, average waiting time, and task completion as the number of completed jobs per run. Second, we report a location-aware operational emissions quantity, aggregated in the implementation as \texttt{total\_ci\_cost\_g} and reported as ``estimated emissions'', obtained by combining the simulated or replayed energy demand with site PUE and time-aligned CI traces. We treat this quantity as a carbon-impact estimate within the evaluated scenario, not as a direct measurement of real-world carbon footprint. This distinction is important since the policy also consumes CI- and PUE-related signals during ranking; accordingly, the reported emissions estimate should be interpreted as an externally recomputed outcome under the scenario model. For each scenario key and repetition, all policies are evaluated on the same generated inputs and arrival schedule; summaries are then aggregated over matched repetitions to compare both absolute outcomes and relative deltas against \texttt{k8s}.

\subsection{Results Analysis}
\label{subsec:results}

\newPolicy consistently reduces the reported operational emissions estimate, at the cost of modest increases in completion time metrics (Figure~\ref{fig:policy-comparison} and Table~\ref{tab:absolute_outcomes}).

Under the evaluated scenarios, \newPolicy, as a reference instantiation within \newFramework, achieves a $45.15\%$ reduction in the reported emissions estimate relative to \texttt{k8s}, substantially larger than the improvements observed for the ranking-based baselines in this subset. In absolute terms, the mean reported operational emissions estimate decreases from $32.14$\,kg to $17.63$\,kg per run (with $\approx 900$ jobs). These results indicate that \newFramework is able to expose and compare policy behaviour in ways that meaningfully reflect edge--cloud heterogeneity in sustainability signals.

The reduction in the reported operational emissions estimate comes with a \textit{controlled} performance overhead: although \newPolicy increases makespan by $1.35\%$, it reduces the average waiting time by $0.04\%$ relative to \texttt{k8s}. Given that edge--cloud scenarios inherently involve locality and resource-fit constraints, this behaviour is consistent with a policy that favours lower-carbon placements when feasible, while maintaining bounded degradation in queueing and completion time metrics. Overall, these results indicate that, under the evaluated hybrid edge--cloud scenarios, \newPolicy reduces the reported operational emissions estimate without substantial degradation in performance metrics.

\subsection{Limitations and future work}
This evaluation is simulation-based and therefore limited by the fidelity of workload models, topology abstractions, and signal assumptions (e.g., CI sampling and mapping). To be more applicable for generic use cases, \newFramework should validate policies on empirical clusters and larger-scale experiments (more sites, higher traffic intensities, and longer horizons), and include additional baselines (e.g., delay-tolerant carbon shifting, carbon-aware bin packing, and network-cost-aware schedulers).

Future work will therefore extend the framework in three directions. First, we will incorporate trace-driven and real-cluster workloads to better reflect workload diversity, including stronger locality constraints, accelerator-heavy jobs, and more variable burst patterns. Second, we will validate the framework on empirical multi-site Kubernetes and federated testbeds using calibrated power measurements, from collected execution traces. Third, we will broaden the sensitivity and robustness analysis across weights, topology settings, exogenous signals, traffic intensity, and workload mixes, to assess how stable policy rankings remain outside the reported slice.

\section{Conclusions}
\label{sec:conclusions_future}

\newFramework demonstrates that a configurable simulation framework can operationalise sustainability-aware scheduling policies for heterogeneous \textit{edge--cloud} settings. In the reported simulator slice, \newPolicy yields a meaningful reduction in the reported operational emissions estimate compared to the default baseline, while incurring only modest overhead in makespan and waiting time. This indicates that the framework is capable of expressing and evaluating the core trade-offs that arise in hybrid deployments.

\section*{Acknowledgements}
The authors acknowledge the funding and support from the Green-DIGIT project ``Greener Future Digital Research Infrastructures'', which has received funding from the European Union’s Horizon Europe research and innovation programme under grant agreement number 101131207.

\bibliographystyle{ACM-Reference-Format}
\bibliography{references}

\appendix

\end{document}